\documentstyle[twoside,fleqn,espcrc2,epsf]{article}

\newcommand{\AmS}{{\protect\the\textfont2
  A\kern-.1667em\lower.5ex\hbox{M}\kern-.125emS}}
\newcommand{\etal}{{\it et al.}}

\def\Dslash{\mathop{\not\!\! D}}

\title{ Tests of Improved Kogut-Susskind Fermion Actions }

\author{K. Orginos and D. Toussaint \address{Department of Physics, University of Arizona, Tucson, AZ 85721, USA}%
        \thanks{Presented by D. Toussaint. Supported by the US DOE.
                Computations were done at ORNL and NERSC as part of 
                the MILC collaboration.}}
       
\begin{document}

\begin{abstract}
Improved Kogut-Susskind quark actions containing repeatedly smeared
links are studied to address the issue of flavor symmetry restoration.
As a measure of the flavor symmetry restoration the mass spectrum of
all the pions is computed.  We present results for several variants of
the ``fat'' actions that are suitable for full QCD simulations.
\end{abstract}

\maketitle

\section{INTRODUCTION}

Improvements of the Kogut-Susskind quark action may allow extraction
of continuum physics from coarser lattices than would be required with
the simplest formulation.  This is especially important for full QCD
simulations, which are much more time consuming than quenched
calculations.  Improvement of the rotational symmetry and the
dispersion relation can be achieved by introducing the Naik term, a
coupling to third nearest neighbors. A more severe problem is flavor
symmetry breaking, which is large for the currently accessible lattice
spacings.  It has been shown that smearing of the gauge links reduces
flavor symmetry breaking. This is because smearing reduces the
coupling to high transverse momentum gluons which cause transitions
among the corners of the Brillouin zone. There are various smearings
proposed so far.  The simplest one, which has been studied by the MILC
collaboration~\cite{MILC_FATLINKS}, is the introduction of the 3-link
staple to the coupling of the nearest neighbors. A more extensive
smearing has been studied by Sinclair and Lagae~\cite{SL}, who
have shown that flavor symmetry breaking is further reduced. Finally,
extensive APE smearing has been studied in SU(2) quenched
spectroscopy~\cite{TD_AH_TK}, showing degeneracy within statistical
errors between the Goldstone pion and the local non-Goldstone pion
($\pi_2$).

 In this paper, we study the effects of several types of smearing on the
flavor symmetry breaking, extending the results reported
in Ref.~\cite{OT}. Our goal is to find an action that achieves
small flavor symmetry breaking, yet is localized enough to
allow for a relatively cheap force computation when used in dynamical
simulations. As a measure of flavor symmetry breaking we use the mass
splittings of all the sixteen pions (in eight separate representations
of the lattice symmetry group)~\cite{GOLTERMAN_MESONS} in
hadron spectroscopy on a common set of stored lattices.

\section{ACTIONS TESTED}

 As a guide in the construction of actions with improved flavor
symmetry breaking we require that the coupling of the
quarks to high transverse momentum gluons is minimized. Consider
 an action which has links smeared by a 3-link staple $S^{(3)}$, a
5-link staple $S^{(5)}$ and a 7-link staple $S^{(7)}$:
\begin{eqnarray}
U_\mu(x)\!\!\!\!&\rightarrow&\!\!\!\! c_1U_\mu(x)+\sum_\nu \Big[ w_3S^{(3)}_{\mu\nu}(x)+\nonumber\\
\!\!\!\!&+&\!\!\!\!\sum_\rho \Big( w_5 S^{(5)}_{\mu\nu\rho}(x) + 
\sum_\sigma w_7 S^{(7)}_{\mu\nu\rho\sigma}(x)\Big)\Big] 
\label{smearing}
\end{eqnarray}
\begin{eqnarray}
\lefteqn{S^{(3)}_{\mu\nu}(x) = U_\nu(x)
         U_\mu(x+\hat\nu)U^\dagger_\nu(x+\hat\mu)}\nonumber\\
\lefteqn{S^{(5)}_{\mu\nu\rho}(x) = U_\nu(x)
         S^{(3)}_{\mu\rho}(x+\hat\nu)U^\dagger_\nu(x+\hat\mu)}\nonumber\\
\lefteqn{S^{(7)}_{\mu\nu\rho\sigma}(x) = U_\nu(x)
         S^{(5)}_{\mu\rho\sigma}(x+\hat\nu)U^\dagger_\nu(x+\hat\mu)}
\label{staples}
\end{eqnarray}

In the weak coupling limit the couplings $V_1,V_2$, and $V_3$ to the
gauge field with one, two or three
of the transverse momentum components $\pm\pi/a$
can be written as functions of the staple couplings $w_3,w_5,w_7$, and
the single link coupling $c_1$:
\begin{eqnarray}
V_1 &=& c_1 +  2 w_3 -  8 w_5 - 48 w_7  \nonumber\\
V_2 &=& c_1 -  2 w_3 -  8 w_5 + 48 w_7  \nonumber\\
V_3 &=& c_1 -  6 w_3 + 24 w_5 - 48 w_7  
\label{vertex}
\end{eqnarray}
The overall normalization condition
\begin{equation}
c_1  + 6 w_3 + 24 w_5 + 48 w_7 = 1
\label{normalization} 
\end{equation}
is used to ensure that the total coupling to the nearest neighbor in
the free field limit is one.  For $c_1 = 2 w_3 = 8 w_5 = 48 w_7 = 1/8
$, all the couplings to gluons with any of the transverse momenta
equal to $\pm\pi$ are zero. This set of parameters defines our
``Fat7'' action.  The ``Fat5'' action is constructed by requiring that
the magnitude of all the couplings $V$ is minimized. The ``Fat5''
couplings are $c_1 = 2 w_3 = 8 w_5 = 1/7$, $w_7=0$, which give
$|V_1|=|V_2|=|V_3| = 1/7 $.

We have also tested an action (``All5'') which contains all the non
self-intersecting length 5 paths that connect nearest neighbors and
third nearest neighbors.  At the free field limit, such an action is
the same as the ``Fat5'' action with a Naik term.  The paths connecting
nearest neighbors divide into three classes. The planar paths that
displace the fundamental link by 0 or 2 sites with total weight $c_1$,
the planar paths that displace the fundamental link by 1 site with
total weight $w_3$, and the non-planar paths with total weight $w_5$.
If we use the ``Fat5'' parameters, appropriately scaled to accommodate
the Naik term, and distribute the weight equally among the members of
each class of paths, the couplings $V$ are minimized.

Together with the MILC collaboration, Anna Hasenfratz and Chet Nieter,
this work is now being extended to include ``APE smeared'' actions,
where the fattened link is projected back on to SU(3). Here we present
two preliminary results. ``Ape1'' has one level of APE smearing with
APE parameter $\alpha=0.75$~\cite{TD_AH_TK}, which was chosen to match
the MILC fat action. The MILC fat action and the ``Ape1'' action
differ only by the projection to SU(3). The
second variation of APE smeared action we tested is ``Ape4'', which
has four APE smearings with $\alpha=0.5$.  

\section{SIMULATIONS AND RESULTS}

For our spectroscopy, we used lattices with dynamical
Kogut-Susskind quarks. These lattices were produced with the Symanzik
improved gauge action (same lattices as those in~\cite{OT}).
The dynamical fermion action used was the MILC 
``fat Naik action'' with Dirac matrix $2m + \Dslash$, where  
\begin{eqnarray}
\Dslash(x,y)&&\!\!\!\!\!\!\!\!=\sum_{\mu=-4,4} \eta_\mu(x)\,sign(\mu)\times\bigg[
\nonumber\\ 
&&\!\!\!\!\!\!\!\!\!\!\!\!\!\!\!\!\!\!\!\bigg(c_1 U_\mu(x)+
                     w_3 \sum_{\nu\neq\mu}S^{(3)}_{\mu\nu}(x)\bigg)\delta_{y,x+\hat\mu}\nonumber\\
&&\!\!\!\!\!\!\!\!\!\!\!\!\!\!\!\!\!\!\!+c_3 U_\mu(x)U_\mu(x+\hat\mu)U_\mu(x+2\hat\mu) \delta_{y,x+3\hat\mu}
\bigg].
\end{eqnarray}
Here, $c_1=(9/8)(1/4)$ is the coefficient of the conventional
single-link term, $c_3=-1/24$ is the coefficient of the third nearest
neighbor (Naik) term, and $w_3=(9/8)(1/8)$ is the coefficient of the
staple term.

Spectroscopy was done on $12^3\times 32$ lattices with
$\beta_{imp}=7.3$ and $16^3\times 48$ lattices with $\beta_{imp}=7.5$.
At $\beta_{imp}=7.3$ we used $m=0.02$ and $0.04$, while at
$\beta_{imp}=7.5$ we used $m=0.015$ and $0.030$.  In order to have a
fair comparison of the actions tested, for each valence action we did
spectroscopy on the same set of lattices and interpolated the spectrum
of all the sixteen pions to the quark mass where $M_G/M_\rho=0.55$
($M_G$ is the Goldstone pion mass).  Successive lattices were
separated by five molecular dynamics time units, and sample sizes
ranged from 48 to 60 lattices.

\begin{figure}[t]
\epsfxsize=7.5cm
\epsfbox{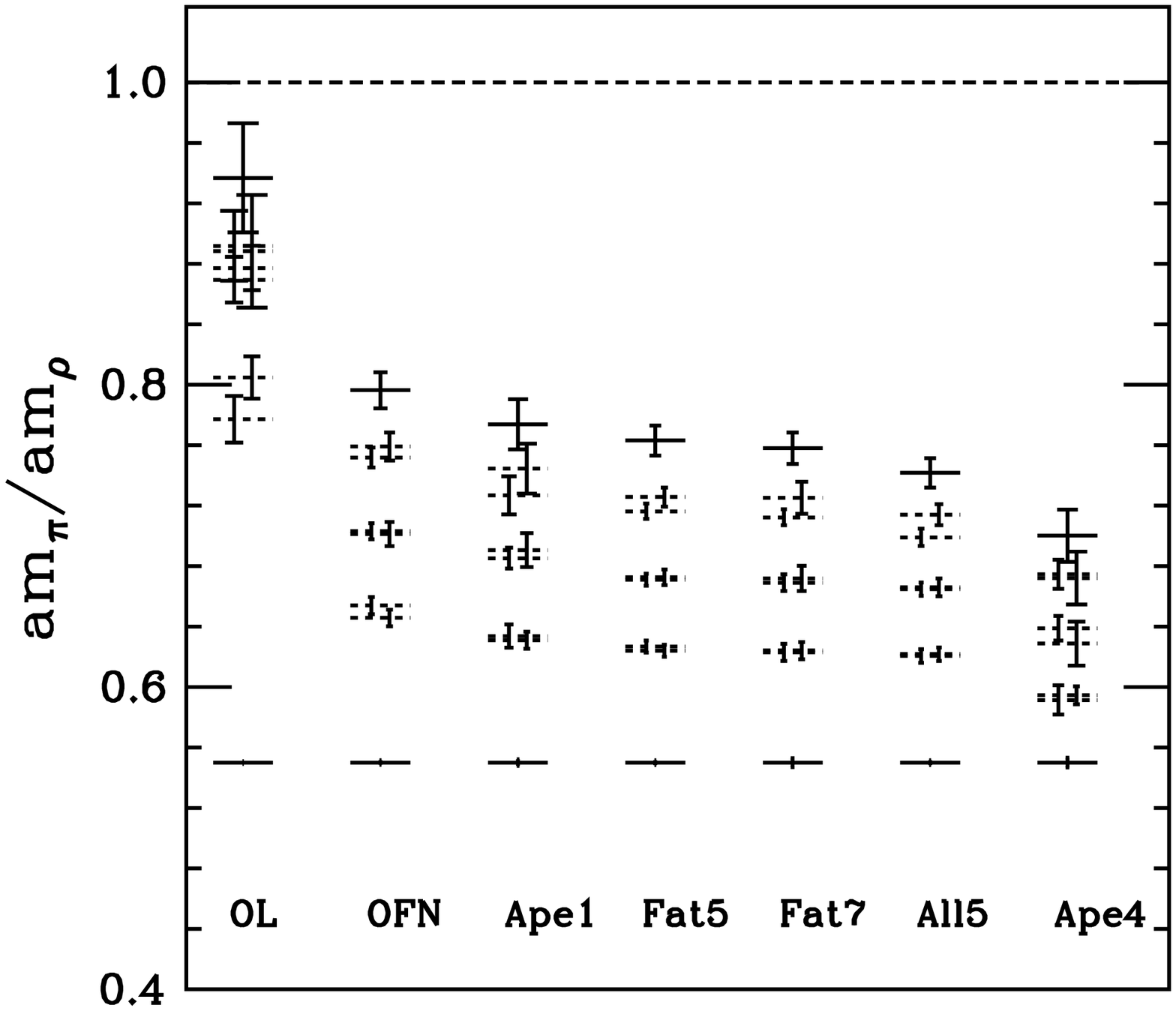}
\vspace{-.2cm}
\caption{Interpolated masses for $\beta_{imp}=7.3$.}
\label{fig:beta73}
\end{figure}
\begin{figure}[t]
\epsfxsize=7.5cm
\epsfbox{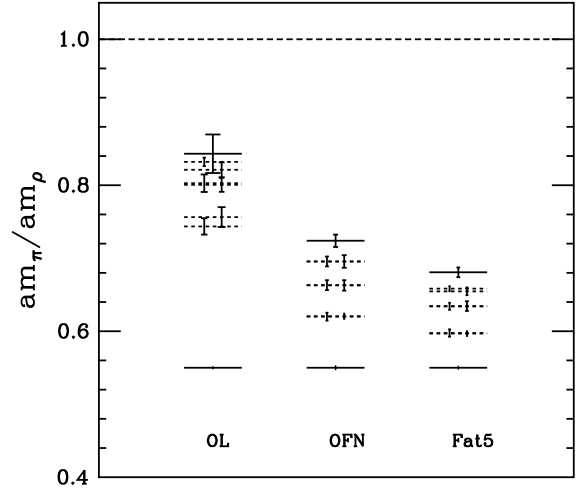}
\vspace{-.2cm}
\caption{Interpolated masses for $\beta_{imp}=7.5$.}
\label{fig:beta75}
\end{figure}

In Fig.~\ref{fig:beta73} and Fig.~\ref{fig:beta75} we present the
spectrum of all the sixteen pions interpolated to $M_G/M_\rho=0.55$,
for $\beta=7.3$ and $\beta=7.5$ respectively.  We present data for the
actions we tested and for comparison we also plot the data for the
standard one link (OL) Kogut-Susskind action and the ``fat Naik
action''(OFN).  The lowest level is always the Goldstone pion at
$0.55$. The highest is the 3-link pion $\gamma_0\gamma_5\otimes{\bf
1}$. The intermediate levels come in nearly degenerate pairs. In order to
distinguish them, they are plotted shifted left and right from the
center. The second level is the $\pi_2$,
$\gamma_0\gamma_5\otimes\gamma_0\gamma_5$ (left) and the 1-link
$\gamma_5\otimes\gamma_i\gamma_5$ (right). The third is the 2-link
$\gamma_5\otimes\gamma_i\gamma_0$ (left) and the 1-link
$\gamma_0\gamma_5\otimes\gamma_i\gamma_j$ (right). The fourth is the
3-link $\gamma_5\otimes\gamma_0$ (left) and the 2-link
$\gamma_0\gamma_5\otimes\gamma_i$ (right). The above degeneracies are
predicted in the contribution~\cite{Sharpe} by Lee and Sharpe in this
conference.


 From our data it can be seen that in general, smearing improves
flavor symmetry breaking. Of the actions tested here, flavor symmetry
breaking is smallest with the ``Ape4'' action, which is also the one
with the most extensive smearing. Unfortunately, such an action is
very costly for dynamical simulations. ``Ape1'' seems better than the
OFN action. This leads us to conclude that the projection to SU(3),
which APE smearing does, contributes to the improvement of the flavor
symmetry breaking. The ``Fat5'' action gives some additional
improvement over the ``Ape1''. It would be interesting to check if a
projection of the ``Fat5'' link to SU(3) would give a further
improvement. The ``Fat7'', which has all the couplings to gluons with
transverse momenta $\pm\pi$, does not improve significantly over the
``Fat5'' action. Finally, ``All5'', which is the same as ``Fat5'' from
the weak coupling point of view, does improve significantly over
``Fat5''.  However, when the cost of the force computation is taken
into account, the best of the above actions for full QCD calculations
may be the ``Fat5'' (with the Naik term added) or even the ``OFN''
action. The decrease in flavor symmetry breaking as $\beta_{imp}$ 
increases form $7.3$ to $7.5$ is consistent with the expected $a^2$
dependence of lattice artifacts with Kogut-Susskind quarks.

Clearly, our data show that one has to look at the spectrum of all
the sixteen pions before drawing any conclusions for the quality of
the action.  In particular, for studies of QCD thermodynamics with the
strange quark~\cite{ADD_STRANGE}, one would like to have all the pions
light compared to the kaons.  
In order to
do so, we have to go to lattice spacings significantly
smaller~\cite{OT} than those predicted from looking just at the local
non-Goldstone pion.

\end{document}